\documentclass[twocolumn,paper,superscriptaddress,aps,prebibtex]{revtex4-2}

\usepackage[utf8]{inputenc}
\usepackage{bbold}
\usepackage{amsmath}
\usepackage{amsfonts}
\usepackage{amssymb}
\usepackage{graphicx}
\usepackage[dvipsnames,svgnames]{xcolor}
\usepackage{xstring}
\usepackage{xparse}
\usepackage{adjustbox}
\usepackage{enumitem}
\usepackage{bbm}
\usepackage{hyperref}
\hypersetup{
	colorlinks=true,
	linkcolor=blue,
	filecolor=magenta,      
	urlcolor=blue,
	citecolor=blue
}

\DeclareMathOperator{\var}{Var}
\begin{document}

\title{Sparse species interactions reproduce abundance correlation patterns in microbial communities}

\author{Jos\'{e}~Camacho-Mateu}
\address{Grupo Interdisciplinar de Sistemas Complejos (GISC), Departamento de Matem\'aticas, Universidad Carlos III de Madrid, Legan\'{e}s, Spain}

\author{Aniello Lampo}
\address{Grupo Interdisciplinar de Sistemas Complejos (GISC), Departamento de Matem\'aticas, Universidad Carlos III de Madrid, Legan\'{e}s, Spain}

\author{Matteo Sireci} 
\address{Departamento de Electromagnetismo y F{\'\i}sica de la Materia, Universidad de Granada, Granada, Spain} 
\address{Instituto Carlos I de F{\'\i}sica Te\'orica y Computacional, Universidad de Granada, Granada, Spain}

\author{Miguel \'{A}ngel Mu\~noz}
\address{Departamento de Electromagnetismo y F{\'\i}sica de la Materia, Universidad de Granada, Granada, Spain} 
\address{Instituto Carlos I de F{\'\i}sica Te\'orica y Computacional, Universidad de Granada, Granada, Spain}

\author{Jos\'{e} A. Cuesta}
\address{Grupo Interdisciplinar de Sistemas Complejos (GISC), Departamento de Matem\'aticas, Universidad Carlos III de Madrid, Legan\'{e}s, Spain}
\address{Instituto de Biocomputaci\'on y F\'isica de Sistemas Complejos (BIFI), Universidad de Zaragoza, Zaragoza, Spain}

\begin{abstract}
During the last decades macroecology has identified broad-scale patterns of abundances and diversity of microbial communities and put forward some potential explanations for them. However, these advances are not paralleled by a full understanding of the dynamical processes behind them. In particular, abundance fluctuations of different species are found to be correlated, both across time and across communities in metagenomic samples. Reproducing such correlations through appropriate population models remains an open challenge. The present paper tackles this problem and points to sparse species interactions as a necessary mechanism to account for them. Specifically, we discuss several possibilities to include interactions in population models and recognize Lotka-Volterra constants as a successful ansatz. For this, we design a Bayesian inference algorithm to extract sets of interaction constants able to reproduce empirical probability distributions of pairwise correlations for diverse biomes. Importantly, the inferred models still reproduce well-known single-species macroecological patterns concerning abundance fluctuations across both species and communities. Endorsed by the agreement with the empirically observed phenomenology, our analyses provide insights on the properties of the networks of microbial interactions, revealing that sparsity is a crucial feature.
\end{abstract}

\maketitle
\begin{figure*}[t!]
  \begin{center}
  \includegraphics[width=0.98\textwidth]{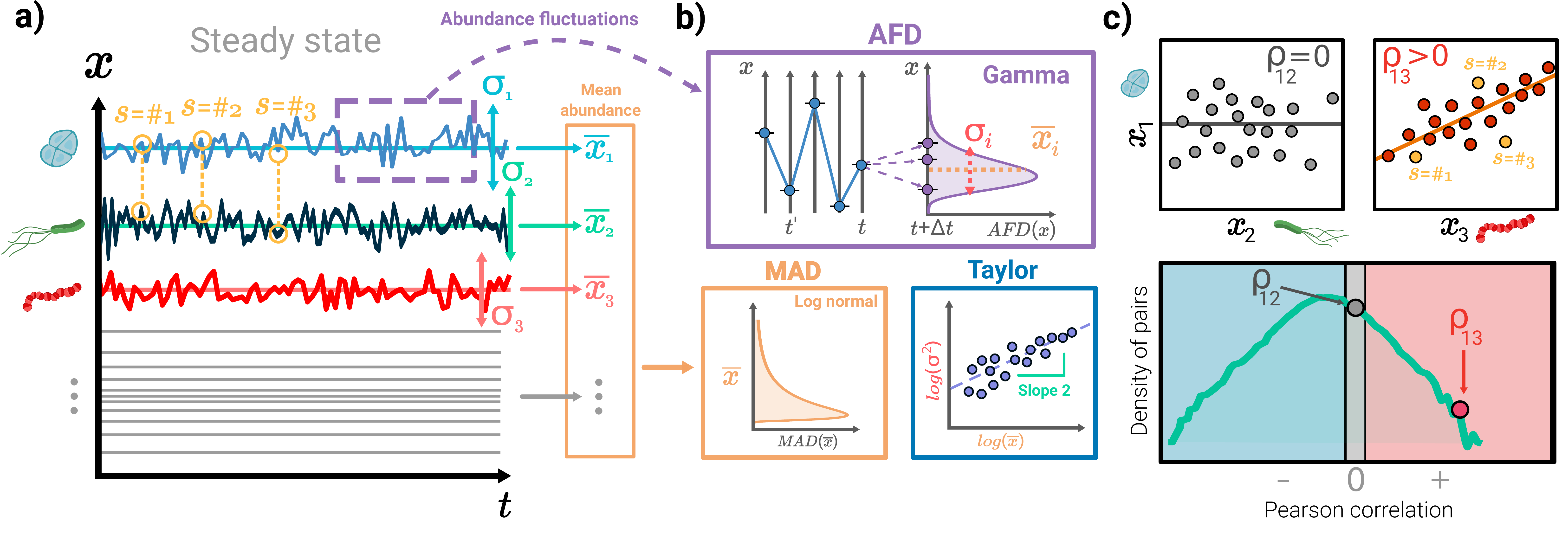}
  \end{center}
  \caption{\label{fig:macroLawsCartoon}
  \textbf{Infographic of the population dynamics and the resulting macroecological patterns.} Panel \textbf{(a)} portrays, as an illustrative example, three individual-species (color coded) time courses at equally spaced times (longitudinal data), resulting from the integration of \eqref{eq:mainEquation}. The abundances at different samples describe the fluctuations around the mean (abundance fluctuation distribution, AFD) are well described by a gamma distribution, as shown in panel \textbf{(b)} (see Figs.~S4 and S5 of the SI). For each species, this distribution is characterized by its mean value $\bar{x}_i$ and its variance $\sigma_i^2$. These two magnitudes are linked by Taylor's law $\sigma_i^2\propto\bar{x}^2_i$ (panel \textbf{(b)}). The mean abundances of all species are distributed as a lognormal (mean abundance distribution, MAD) (panel \textbf{(b)}). Further details about Taylor's law and MAD are presented in Figs.~S6 and S7 of the SI. Panel \textbf{(c)} illustrates the correlations between abundance fluctuations of pairs of species across samples (a point for each sample/realization). The top-left plot illustrates the case of two uncorrelated species whereas the top-right plot illustrates two positively correlated species. The bottom picture shows the distribution of Pearson's coefficients $\rho_{ij}$
  %(cf.~\eqref{eq:PearsonCoefficient})
  of all pairs of species. Empirically, this distribution is found to generally cover the entire range $-1\le\rho_{ij}\le 1$ and to exhibit a peak at negative values.}
\end{figure*}

\section{Introduction}

Our understanding of the microscopic living world has been recently challenged by the advent of metagenomics \cite{Handelsman1998,Steele2005}. Indeed, DNA sequencing methods unveiled that a large fraction of microbial diversity was missing in laboratory cultures \cite{Whitman1998,Rappe2003,Hug2016ANV}. Moreover, the possibility to collect genetic material directly from its natural environment introduced a new dimension---the set of samples---along which the properties of the biome may vary. This has given rise to the production of the largest datasets ever, allowing microbial communities to be investigated at a much greater scale and detail than before. 

To approach this new profusion of data, macroecology---the quantitative analysis of emergent broad-scale patterns---prevailed as a promising point of view \cite{brown1989macroecology,brown1995macroecology,field2002macroecology,prosser2007role,Shade2018,McGill2018}. The framework paved the way to assess statistically the variation in abundance and diversity that, despite the complexity of the underlying microscopic behaviors, often portray distinctive distributions and that sometimes may be explained in terms of basic ecological forces. Specifically, considerable progress has been achieved in the observation of statistical regularities of taxa populations across time \cite{Ji2020}, spatial samples \cite{Zaoli2021}, and species-abundance distributions \cite{Shoemaker2017}. 

Most remarkably, a recent paper by J. Grilli \cite{Grilli2020} provided an important step towards a macroecological study of microbial communities. Relying on the analysis of data from nine real biomes, the work characterizes some patterns of abundance variation in terms of (using the terminology of e.g.~\cite{Ji2020,Shoemaker2017,Grilli2020}) three macroecological laws (see Fig.~\ref{fig:macroLawsCartoon}): \textit{i}) the fluctuations in the abundance of any given species across samples follow a gamma distribution; \textit{ii}) the variances of these distributions for different species are proportional to the square of their means (a particular case of Taylor's law \cite{Taylor1961} for power-law exponent $2$); and \textit{iii}) the mean abundances across species follow a lognormal distribution. These macroecological patterns of species fluctuations and diversity have been parsimoniously explained using the Stochastic Logistic Model (SLM), which  endows the traditional logistic equation with a (multiplicative) stochastic term \cite{MN1,MN2} embodying information about environmental variability \cite{Grilli2020,Descheemaeker2020,Wolff2023,Shoemaker2023}.

Beside the aforementioned patterns, the analysis of empirical data unveils also the existence of non-trivial pairwise correlations in species abundances \cite{Grilli2020}. In particular, the Pearson's correlation coefficients of all pairs of species in a biome display distributions ranging from anti-correlations to positive correlations, with a peak often located at negative values (see Table~S1 of the SI). These pairwise correlations are not accounted for by the SLM model because it treats species dynamics as independent from each other \cite{Sireci2023}. Describing correlations in species abundances calls, thus, for introducing some sort of interaction between species. 

The existence of species interactions in microbiomes is well documented in a wealth of experimental results that manage to observe and measure them \cite{Kehe2021, Hu2022, Shetty2022a, Weiss2022}. Indeed, microbial interactions are a key ingredient behind community stability \cite{Coyte2015, Butler2018}, necessary for, e.g., the maintenance of health in human biomes \cite{Lozupone2012, Relman2012, DeDios2017} or the control of medical disorders \cite{Newell2014, Khanna2017, Frank2007, Palmer2022}. From the modeling standpoint, microbial pairwise interactions have also received a lot of attention in the field of network inference, where researchers struggle with the problem of reconstructing species-interaction networks from available empirical datasets \cite{Xiao2017, Angulo2019, Matchado2021, Pinto2022}. All these efforts make evident the current consensus on the crucial role of interactions in microbial ecosystems, justifying their inclusion in modeling approaches.

Interactions can be implemented in models in at least two ways: (i) \emph{indirectly,} i.e. assuming the diverse environmental noise terms to be correlated with each other or (ii) \emph{directly,} i.e. introducing a coupling  between species abundances, or (iii) using a combination of both. The first route assumes that correlations in the abundance of two species arise from similar or opposite responses of both species to changes in the environment (variation of nutrients, presence of chemicals, changes in temperature or pH, etc.). As a matter of fact, this strategy has recently been explored in connection with phylogeny, under the rationale that genetically related species tend to respond to environmental cues alike \cite{Sireci2023}. Coupling the noise terms has the added advantage of being a modification of the SLM that preserves, by construction, Grilli's three empirical laws. However, as we will later show, environmental noise by itself is insufficient to fully capture the correlation patterns observed empirically so that it needs to be replaced---or at least supplemented---by the inclusion of direct species interactions.

In this paper, we propose a population model which includes direct Lotka-Volterra pairwise couplings between species abundances. This approach is suitable to model competition mechanisms (negative correlations) detected in real biomes and their interplay with cooperative ones, as well as other kinds of relationships. Unlike other approaches we do not attempt to infer specific pairwise interactions but, rather, the ensemble of possible interaction networks able to reproduce the empirically observed correlation patterns. We will show that interactions provide a necessary and sufficient requirement to account for the observed correlation distributions, besides preserving Grilli's three empirical laws. Our analysis identifies sparsity, \textit{i.e.} the low density of species interactions, as a critical feature of microbial networks. This feature suggests a prevalence of amensalistic and commensalistic relationships among the community biota.

\begin{figure}
  \begin{center}
 \includegraphics[width=0.99\columnwidth]{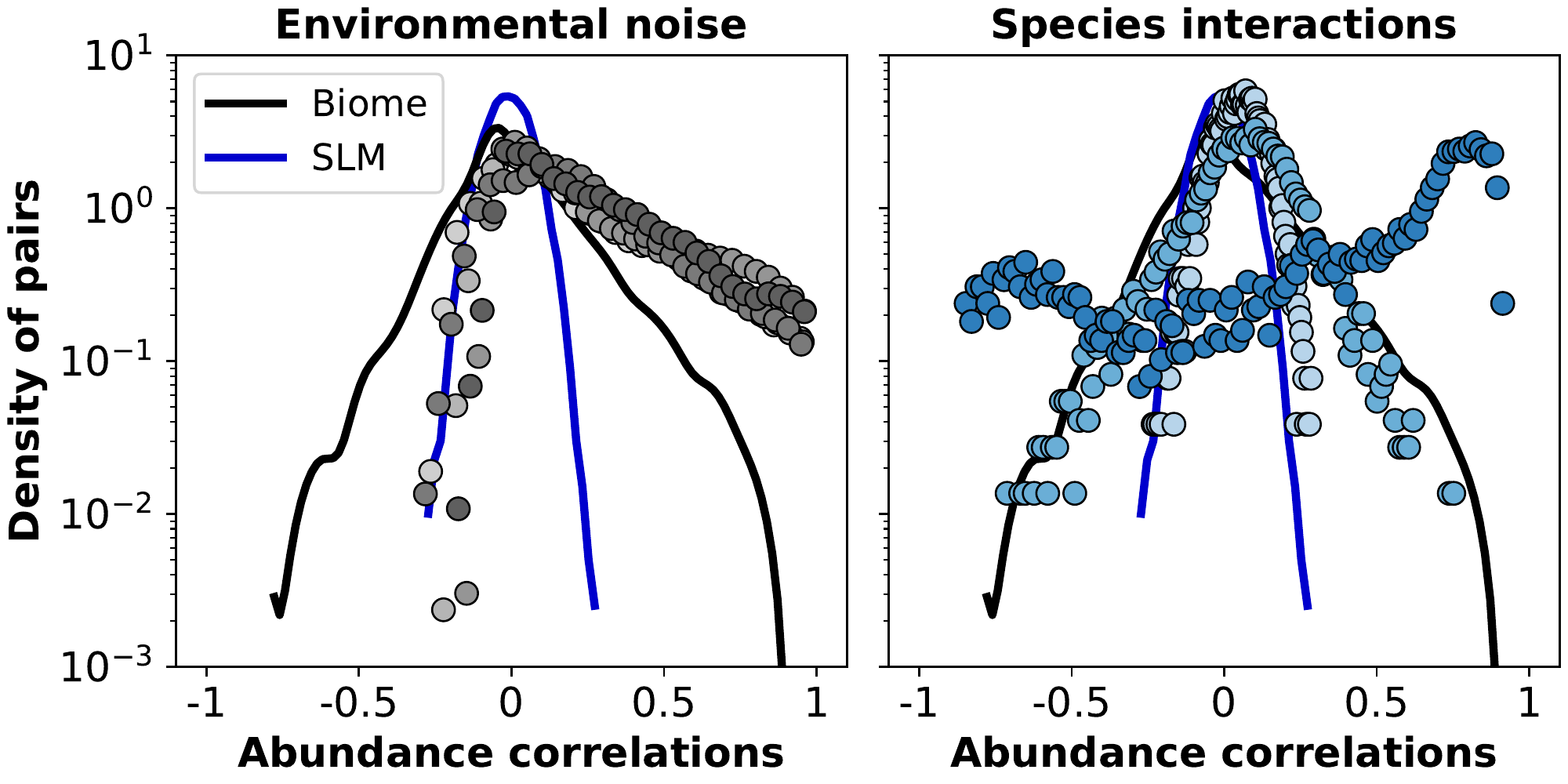}
  \end{center}
  \caption{\label{fig:CorrelationSampling}
Distributions of Pearson's abundance correlation coefficients as obtained in the model with  (left panel) a few samples of the noise correlation matrix $\mathbf{W}$ (each with a different gray shade) or (right panel)  with random samples of the Lotka-Volterra matrix $\mathbf{A}$. The black solid lines portray in each case the empirical distribution as obtained from the \textit{Seawater} microbiome (species which appear in less than 50\% of the communities have been filtered out), while the blue ones represent the distribution of correlations as obtained from the  model without interactions. In the left plot, colored circles show the results for a few samples of matrices $\mathbf{W}$ (see `Material and methods' for details of the sampling procedure); Lotka-Volterra constants are chosen as $a_{ij}=-\delta_{ij} / K_i$, with carrying capacities $K_i$ sampled from a lognormal distribution with mean $0.1$ and standard deviation $0.5$---as for the SLM \cite{Grilli2020}. The results shown in this figure are typical (see Secs.~7B and C of the SI for a more thorough exploration). In the right plot, colored circles represent correlations resulting from the SLVM with $\mathbf{W}=w\mathbf{I}$ and Lotka-Volterra constants $a_{ij}$ ($i\ne j$) sampled from a Gaussian distribution with zero mean and standard deviation $0.03$. A random selection of $60\%$ of such constants are set to zero (i.e.~the connectance of the interaction matrix is $C=0.4$).}
\end{figure}

\section*{Modeling microbiomes}

\begin{figure*}[t!]
  \begin{center}
  \includegraphics[width=0.99\textwidth]{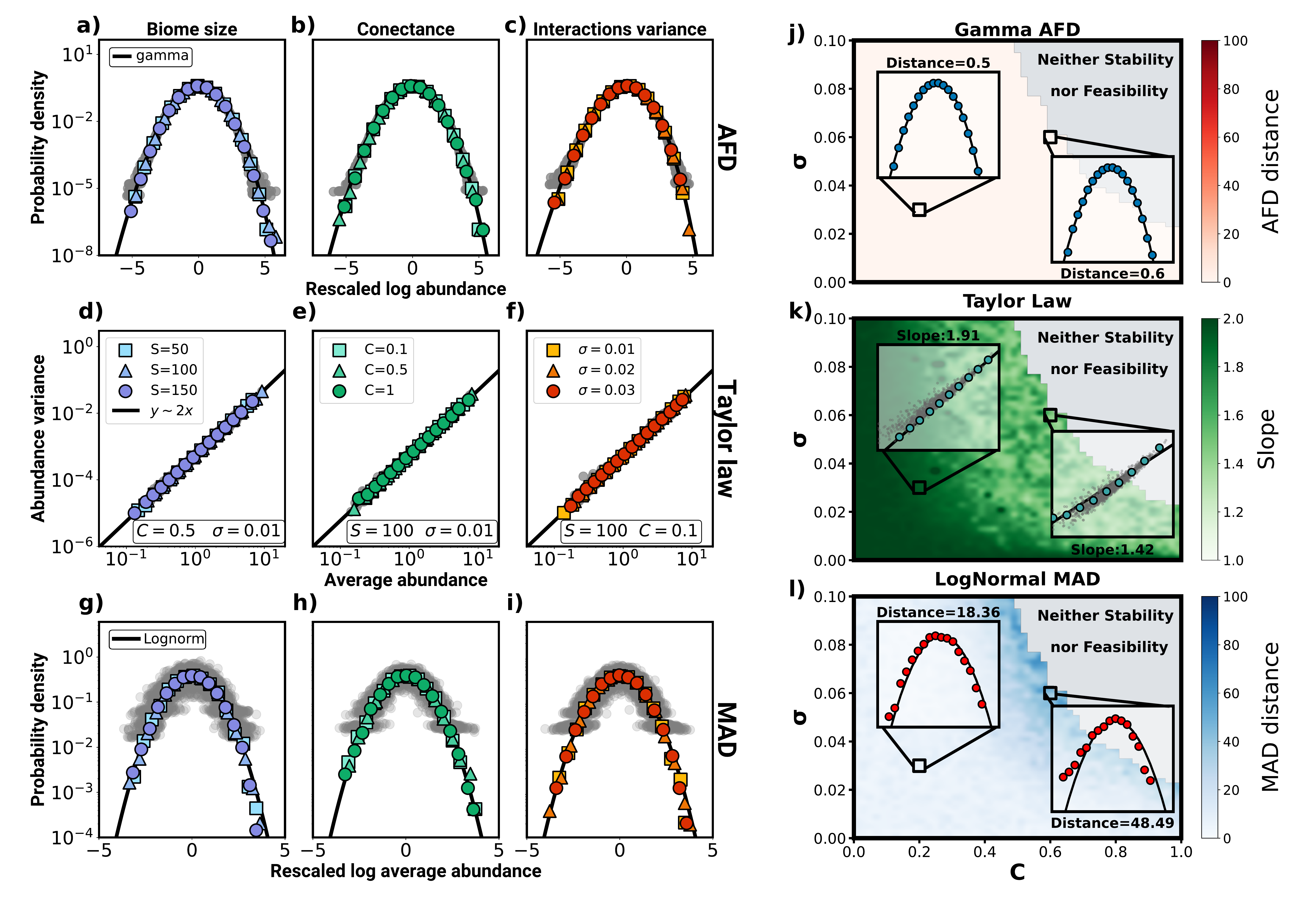}
  \end{center}
  \caption{\label{fig:macroLawsPlot}
Grilli's three macroecological laws as a function of the interaction parameters. Specifically, the figure shows the abundance fluctuation distribution (AFD) (panels \textbf{(a)}--\textbf{(c)}), Taylor's law (panels \textbf{(d)}--\textbf{(f)}) and the mean-abundance distribution (MAD, panels \textbf{(g)}--\textbf{(i)}) for different values of the species number $S$ (panels \textbf{(a)}, \textbf{(d)}, \textbf{(g)}), the connectance $C$ (panels \textbf{(b)}, \textbf{(e)}, \textbf{(h)}), and standard deviation of the interaction constants $\sigma$ (panels \textbf{(c)}, \textbf{(f)}, \textbf{(i)}). Results have been averaged over $100$ realizations of the SLVM (\eqref{eq:mainEquation}) each one with a different random interaction matrix. Results including all realizations are depicted as a cloud of gray points, whereas averages are shown as colored bullets. The AFD obtained for a given realization contains the results for all species, represented in terms of rescaled logarithm abundances ($z=\var(x)^{-1/2}\log(x/\bar{x})$). Solid black lines correspond to gamma distributions. MAD plots \textbf{(g)}--\textbf{(i)} is obtained by properly rescaling the mean abundances, and are fitted by a normalized (zero mean, unit standard deviation) lognormal distribution (black solid line). Similarly, the black straight lines in panels \textbf{(d)}--\textbf{(f)} describe the relation $\var(x_i)\propto\bar{x}_i^2$ in logarithmic scale. Panels: \textbf{(j)}, \textbf{(k)}, \textbf{(l)} illustrate the limits of the weak-interaction regime across the set of parameters that characterize species interactions. The plots quantify the compliance with \textbf{(j)} a gamma AFD, \textbf{(k)} Taylor's law, and \textbf{(l)} a lognormal MAD, within the region where the system is stable and feasible. Each pixel corresponds to a combination of values of the network connectance $C$ (horizontal axis) and the standard deviation $\sigma$ of the distribution of interactions (vertical axis). The color of the pixel quantifies the distance from the AFD to a gamma distribution \textbf{(j)}, the value of the exponent $\gamma$ in the relationship $\var(x_i)\propto\bar{x}_i^{\gamma}$ \textbf{(k)}, and  the distance of the MAD to a lognormal distribution \textbf{(l)}, averaged over a sample of $100$ realizations. Gray areas mark the region of the parameter space where the resulting systems are neither stable nor feasible. In these plots $S=50$, $\tau_i=0.1$, $w=0.1$, and the carrying capacities are sampled from a lognormal distribution (mean $0.1$, standard deviation $0.5$).}
\end{figure*}

\subsection*{Environmental noise vs.~species interactions}

A simple model that couples species in a parsimonious way 
is the Stochastic Lotka-Volterra Model (SLVM)
\begin{equation}\label{eq:mainEquation}
    \dot{x}_i=\frac{x_i}{\tau_i} \left(1+\sum_{j=1}^S a_{ij} x_j\right)+x_i \xi_i, \quad i=1,\dots,S,
\end{equation}
where $x_i(t)$ is the abundance of species $i$ at time $t$, $\tau_i$ is the time scale of its basal population growth, and $\xi_i$ is a zero-mean, multivariate Gaussian white noise (It\^o interpretation) with correlations $\langle\xi_i(t)\xi_j(t^\prime)\rangle= w_{ij}\delta(t-t^\prime)$. The matrix $\mathbf{W}=(w_{ij})$ accounts for environmental fluctuations, whereas the off-diagonal terms of matrix $\mathbf{A}=(a_{ij})$ describe direct, Lotka-Volterra-like interactions between species, and the diagonal terms $a_{ii}=-1/K_i$ incorporate the carrying capacity for each species $i$.

When $\mathbf{W}=w\mathbf{I}$ and $\mathbf{A}$ is a diagonal matrix, \eqref{eq:mainEquation} becomes the SLM. By turning on the off-diagonal terms of the noise correlation matrix $\mathbf{W}$ (indirect interactions) and/or of the Lotka-Volterra matrix $\mathbf{A}$ (direct interactions), we can study the effect of correlated environmental noise and/or direct species interactions on species pairwise correlations. Ideally, though, the model should contain the right proportion of both terms.

Before proceeding, let us remark that both coupling terms $\mathbf{A}$ and $\mathbf{W}$ could have been derived, after some simplifying assumptions, from a more complex consumer-resource type of model describing explicitly the dynamics of environmental factors or resources (see e.g.~\cite{Ho2022,Sireci2023} and references therein). However, for the sake of simplicity and to allow for a straightforward comparison with the SLM, here we build our model by proposing general forms for these two types of couplings independently.

At first sight, adding environmental noise correlations ($\mathbf{W}$) has an advantage over adding interactions ($\mathbf{A}$) in that Grilli's first law is preserved by construction (see Sec.~7A of the SI). The second law simply amounts to setting $w_{ii}=w$ for all $i$. As for the third law, it can be fulfilled if one chooses ad hoc the carrying capacities $K_i$ as lognormal distributed random variables \cite{Grilli2020}. Obviously, these latter choices do not explain the origin of the second and third laws, but at least render a model that is compatible with them. On the downside though, the fact that $\mathbf{W}$, being a covariance matrix, must be symmetric and positive definite severely constrains the kind of abundance correlations that \eqref{eq:mainEquation} can generate.

On the other hand, if one introduces interactions while keeping $\mathbf{W}=w\mathbf{I}$, in general the first and second laws do not hold exactly---although they may approximately do so. However, the presence of interactions strongly affects the average abundance of the species. While the SLM (with or without a noise correlation matrix) predicts a stationary population that fluctuates around its carrying capacity, in the presence of coupling, the mean values are the solution of the linear system (see Sec.~2 of the SI)
\begin{equation}\label{eq:shifts}
    \sum_{j=1}^Sa_{ij}\bar{x}_j=\frac{\tau_iw}{2}-1, \quad i=1,\dots,S,
\end{equation}
where $\bar{x}_j$ denotes the average abundance of species $j$. Therefore interactions shift these average abundances to the extent that, even if all carrying capacities were the same, the $\bar{x}_j$ would split over a range of values. This may not be a full explanation of the third law yet, but it opens the possibility that its origin might lie on a particular structure of the network of interactions. As a matter of fact, Descheemaeker et al. \cite{Descheemaeker2021} were able to obtain a lognormal mean abundance distribution by simply introducing an indirect interaction between species through a global carrying capacity of the system.

 A quick test to decide which of the previously described two strategies is most promising to model abundance correlations is to generate a large sample of random matrices (either $\mathbf{W}$ or $\mathbf{A}$), and for each of them simulate the stochastic process~\eqref{eq:mainEquation}, calculate abundance correlations between pairs of species, and compare the resulting distributions with those empirically obtained from the microbiome datasets with the same number of species (Fig.~\ref{fig:macroLawsCartoon}\textbf{a} and \textbf{c}). Each of these two samples must fulfill some constraints: matrices $\mathbf{W}$ must all be symmetric and positive definite, and matrices $\mathbf{A}$ must all lead to a \emph{feasible} (i.e.~$\bar{x}_i>0$ for all $i$) \cite{Grilli2017} and \emph{asymptotically stable} (i.e.~small perturbations must die out \cite{May1972,Allesina2012}) steady state (see Sec.\ 3 of the SI).

Figure~\ref{fig:CorrelationSampling} shows the distribution of Pearson's abundance correlation coefficients for all $S(S-1)$ pairs of species $i\ne j$, as obtained from a typical dataset and using each of these two matrix ensembles. The empirical distribution decays exponentially to the left and to the right, is a bit asymmetrical (skewness coefficients close to or larger than $1$ for all biomes; see Table~S1 of the SI), and has a peak at slightly negative values of the correlation. The distributions obtained from the $\mathbf{W}$ samples bear little resemblance to the former---they exhibit very little negative correlations, are strongly asymmetric, and show a peak at zero. On the contrary, distributions obtained from the $\mathbf{A}$ samples have a wide range of sample to sample variability, and some of the realizations are very similar to empirical data, often peaking at negative correlation values.

These analyses suggest that environmental noise by itself seems incapable of generating correlations resembling those observed in real microbiomes, and so interactions have to be included in the model. In any case, the presence of correlated noise cannot be ruled out from these analyses, but in order to keep things simple, we henceforth take $\mathbf{W}=w\mathbf{I}$ and focus on the effect of interactions in the model.

\subsection*{Grilli's laws in the presence of interactions}

The model described by \eqref{eq:mainEquation} with $\mathbf{W}=w\mathbf{I}$ and a nontrivial interaction matrix $\mathbf{A}$ is not guaranteed to satisfy any of the three macroecological laws found by Grilli \cite{Grilli2020}---even if the carrying capacities $K_i$ are sampled from a lognormal distribution, as in the SLM. However, as long as the interactions are a `small' perturbation to the SLM, one can reasonably expect them to hold, at least approximately. In particular, if one sets all off-diagonal interaction coefficients to zero, except for a  fraction $C$ (`connectance') that are randomly and independently drawn from a Gaussian distribution $\mathcal{N}(0,\sigma)$, a criterion for the weakness of interactions is that the resulting system remains feasible and asymptotic stable; in other words, $\sigma \sqrt{SC} K_{\text{max}}\ll 1$, where $K_{\text{max}}$ is the maximum carrying capacity (see Sec.~4 of the SI). We will refer to this as the `weak-interaction regime'. 

Figure~\ref{fig:macroLawsPlot} shows the compliance with the three macroecological laws for different combinations of parameters within the weak-interaction regime (see Fig.~\ref{fig:macroLawsCartoon}\textbf{a} and \textbf{b} for the sampling procedure). The first row illustrates that fluctuations of the abundance around the mean values still follow a gamma distribution (first law); the second row reveals that $\var(x_i)\propto\bar{x}_i^2$, according to the second law; and the third row shows that the mean abundances very closely follow a lognormal distribution (third law).  Particularly noteworthy is the compliance with the third law, given that the mean abundances are no longer fixed by the carrying capacities (see \eqref{eq:shifts}), which do follow a lognormal distribution.

Importantly, the gamma abundance-fluctuation distribution (AFD) remains unaffected regardless of the values taken by the interaction parameters (Fig.~\ref{fig:macroLawsPlot}\textbf{j}). Moving closer to the boundary of the weak-interaction regime we can see that Taylor's law still holds, but the exponent gets modified as $\var(x_i)\propto\bar{x}_i^{\gamma}$. As this boundary is approached, the exponent decreases down to values around $\gamma\approx 1.4$ (Fig.~\ref{fig:macroLawsPlot}\textbf{k}) and, likewise, the mean-square distance between the distribution of mean abundances and a lognormal increases (Fig.~\ref{fig:macroLawsPlot}\textbf{l}), although it is never very large.

It is worth mentioning that the SLVM (\eqref{eq:mainEquation}) provides an alternative way to comply with the third law other than sampling the carrying capacities from a lognormal distribution and remaining in the weak-interaction regime. Even if we choose constant carrying capacities ($K_i=K$ for all $i$), \eqref{eq:shifts} allows us to seek interaction matrices that shift the mean abundances so as to follow a lognormal distribution. Section~5 of the SI shows that such matrices do actually exist and yield stable and feasible communities. This finding brings species interactions in the long debate about the origin of heavy-tailed abundance distributions, something which, to the best of our knowledge, has been scarcely investigated (but see \cite{Descheemaeker2020}). This is an issue that goes beyond the aim of the current work and will be explored in a forthcoming publication. Therefore, hereafter we focus on the weak-interaction regime with log-normally distributed carrying capacities. 

\subsection*{Interactions reproduce the distribution of correlations}

\begin{figure*}[t!]
  \begin{center}
 \includegraphics[width=0.99\textwidth]{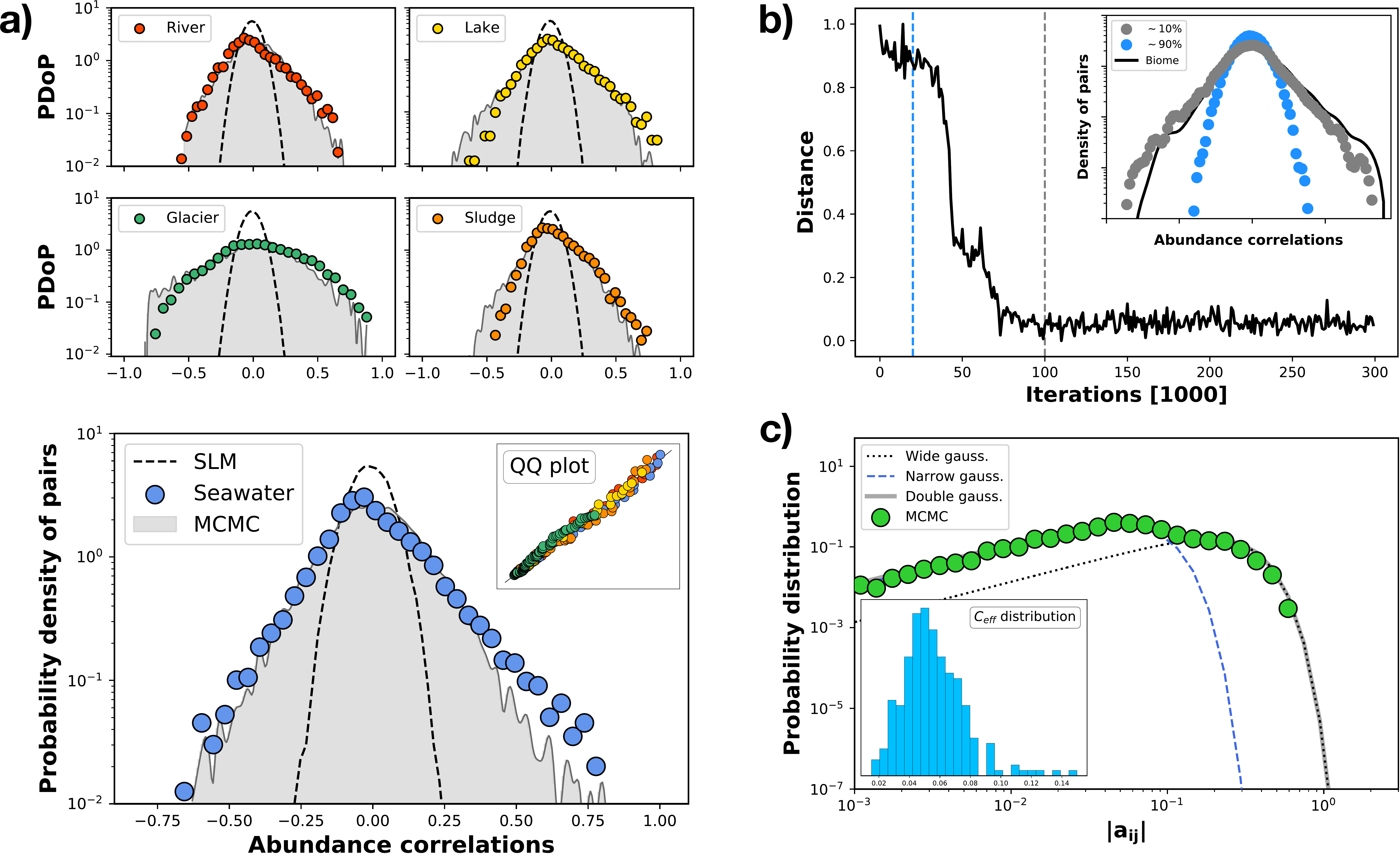}
  \end{center}
  \caption{\label{fig:correlationDistribution}
Abundance correlation distributions for real and simulated communities. In \textbf{(a)}, different colored bullets correspond to different biomes selected from the \textit{EBI metagenomics} platform \cite{Mitchell2018} (namely Seawater, River, Lake, Glacier, and Sludge communities). Black dashed lines portray the distribution of Pearson's coefficients for the abundance correlation of all pairs of species resulting from the SLM. Gray curves show the same distributions as obtained from the SLVM (c.f.~\eqref{eq:mainEquation}), with the Lotka-Volterra interaction constants inferred using the Bayesian approach described in `Material and methods'. The inset in the bottom panel of \textbf{(a)} presents the quantile-quantile (QQ) plot, comparing quantiles of the empirical and the synthetic distributions, for the different biomes. The dots sit on the bisector line, indicating the close alignment between the quantiles of both distributions for each single biome. This test is consistent with the results of a Kolmogorov-Smirnov test ($p$-values of Seawater and River: $0.99$; Lake and Glacier: $0.90$; Sludge: $0.81$). The top panel of \textbf{(b)} shows the Euclidean distance between the logarithms of the empirical and the synthetic distributions (for \textit{Seawater,} using only species appearing in at least 50\% of the samples) vs. the iterations of the MCMC. In blue and grey dots, the inset shows the synthetic distributions obtained at the iteration marked by a dashed line of the corresponding color. The empirical distribution is drawn with a solid black line. \textbf{(c)} shows the distribution of absolute values of the interaction constants ($|a_{ij}|$) for a collection of over $200$ matrices generated through the MCMC method (green bullets). This distribution can be fitted by a convex combination of two Gaussian distributions (grey solid line). The black (blue) dashed line fits the broader (narrower) Gaussian. Practically all coefficients in the narrower Gaussian are negligible compared to the broader one. Hence, the presence of a small fraction of large coefficients gives rise to an effective connectance ($C_{\text{eff}}$) in the associated network. The inset shows a histogram of the values of $C_{\text{eff}}$. It peaks around $0.05$, with tails extending approximately in the range $0.01\lesssim C_{\text{eff}}\lesssim 1.4$. This reveals the high sparsity of the interactions.}
\end{figure*}

An analysis of empirical data selected from the \textit{EBI metagenomics} platform \cite{Mitchell2018} reveals that, on top of the three single-species macroecological laws that we have discussed so far, microbiomes exhibit pairwise correlations. As a matter of fact, the distribution of all $S(S-1)$ Pearson's coefficients of a microbial community has a characteristic pattern (Fig.~\ref{fig:macroLawsCartoon}\textbf{c}). For all the microbiomes that we have considered, this distribution approximately covers the whole range of values ($-1\le\rho_{ij}\le 1$), and is very different from the residual narrow distribution peaked at zero that results from the approach with no species interactions \cite{Grilli2020, Wolff2023, Descheemaeker2020} (see Fig.~\ref{fig:CorrelationSampling}, as well as Fig.~\ref{fig:correlationDistribution}\textbf{a}). Worth noticing is the almost exponential decay to both sides of the interval, and the location of the maximum at slightly negative values of $\rho_{ij}$. 

In order to find a set of interaction matrices $\mathbf{A}$ (or ``ensemble'') that are capable of inducing the empirically observed patterns of correlations, while at the same time preserving Grilli's three laws, we have adopted a Bayesian approach. We know from the previous analysis (Fig.~\ref{fig:CorrelationSampling} right) that matrices inducing similar correlations do exist. Thus, we take the empirical correlation distributions, as well as Grilli's second and third laws, as given---within a Gaussian error---and wonder about the posterior probability distribution of interaction matrices $\mathbf{A}$. Needless to say, this distribution cannot be computed analytically, so in order to sample matrices $\mathbf{A}$ out of the ensemble of possible solutions we need to perform a Markov-chain Monte Carlo (MCMC) simulation (see `Materials and methods' for the details).

As an illustration of the results of this approach, Fig.~\ref{fig:correlationDistribution}\textbf{a} shows the distribution of Pearson's coefficients obtained for five biomes (Fig.~S15 of the SI contains the results for all available biomes in our dataset), along with the empirical ones. The figure speaks for itself as the agreement is rather precise in all cases. Remarkably, the interaction matrices $\mathbf{A}$ which this method converges to follow an interesting statistical pattern. Most coefficients remain zero, and the nonzero ones are distributed as a combination of two zero mean, Gaussian distributions with standard deviations differing from each other in more than one order of magnitude (see Fig.~\ref{fig:correlationDistribution}\textbf{b} for a typical fit). This yields matrices that are highly sparse (sparsity is estimated from the contribution of the wider Gaussian; see `Materials and methods').

It is worth mentioning that removing the constraint on the second and third laws, results similar to those of Fig.~\ref{fig:correlationDistribution}\textbf{a} can be obtained with more densely connected matrices. However, the exponents of Taylor's law are smaller than $2$, and the MAD deviates from a lognormal distribution. This strongly suggests that loosely connected interaction networks (sparsity) might be an important feature of real microbiomes---something that seems to be consistent with existing experimental evidence \cite{Kehe2021,Weiss2022}.

\section*{Discussion}

The existence of macroecological, one-species, statistical patterns in microbial communities puts strong constraints to their mathematical modeling. A remarkable result is that a simple model that neglects interactions between species (the SLM \cite{Grilli2020,Descheemaeker2020,Wolff2023}) seems enough to reproduce such one-species patterns. This is somehow unsettling because species interactions are well-documented to play a fundamental role in the behavior of microbial communities. As a matter of fact, interactions may underlie critical features associated with, e.g., health disorders \cite{Newell2014} such as Crohn’s disease \cite{Khanna2017} or some forms of inflammatory bowel syndrome \cite{Frank2007}, and many current treatments rely on competition among  bacteria \cite{DeDios2017,Palmer2022}.

 However, not surprisingly, neither the SLM nor any other independent-species approach is able to account for the non-trivial patterns of species-abundance correlations that microbial communities exhibit. It is true that the existence of correlation between a pair of species does not necessarily imply a direct interaction between them: it may be caused by  common (similar or opposite) responses to environmental fluctuations or external driving forces. As a matter of fact, phylogenetic closeness can justify why some species respond to the same chemicals alike, which may account for much of the positive correlation observed \cite{Sireci2023}. However, the widespread presence of negative correlations renders this explanation incomplete.

We have shown in this paper that environmental fluctuations by themselves are insufficient to reproduce the correlation patterns observed in real microbiomes, leaving direct interactions as a necessary ingredient in any sound dynamical model of microbial communities. This conclusion is also supported by a recent work \cite{Ho2022} where microbial behavior has been investigated within the framework of the consumer-resource model. In particular, the authors argue that resource competition could account for numerous statistical patterns observed in abundance fluctuations across diverse microbiotas, encompassing the human gut, saliva, vagina, mouse gut, and rice. Significantly, they also investigate the distribution of abundance correlations and find that the predictions derived from considering resource competition approach much closer to empirical data than those related to non-interacting models.

Furthermore, we have shown here that it is possible to introduce interactions to the SLM, thus generating a Lotka-Volterra type of model, such that it complies simultaneously with the three single-species macroecological laws put forward in \cite{Grilli2020} as well as with the abundance-correlation patterns. Interaction matrices satisfying such constraints have been generated through a Bayesian approach implemented by means of a Markov Chain Monte Carlo method. These matrices (and their associated networks) are assumed to be representative of a larger ensemble, whose characterization in terms of topological or structural features is challenging.

We have found very robust evidence that these matrices need to be sparse, which is in agreement with the reported sparse nature of microbial interaction networks \cite{Kehe2021,Weiss2022}. Network sparsity has been previously argued to be a crucial topological feature for the functionality of ecological (and other) networks. For instance, Busiello et al. \cite{busiello2017explorability} have shown that network sparsity is an emergent property resulting from the conflicting forces of optimizing both \emph{explorability} and \emph{dynamical robustness,} where explorability is a measure of the system's ability to adapt to newly intervening changes, and dynamical robustness is the capacity of the system to remain stable after perturbations of the underlying dynamics. In other words, networks able to adapt in a flexible way to external changes, while keeping a robust dynamical regime, need to be sparse.

The high sparsity values that we obtain imply that more than $90\%$ of the pairwise interactions emerging from our analysis are commensalistic or amensalistic (see Fig.\ S20). This is an interesting outcome because in microbiotas, commensalism can be associated to cross-feeding and amensalism to poisoning by toxins, two types of interactions that are very common between microbes \cite{DSouza2018,Goldford2018}, and are known to help stabilizing diverse ecological communities \cite{Mougi2016}.

Beside network sparsity we have not been able so far not identify other structural features telling this ensemble of networks apart from random ones, in the same way as previous work distilled properties, such as nestedness, modularity, or trophic-level (hierarchical) organization, from the analyses of other (much smaller and less complex) ecological networks.

For example, the relative frequency of different types of interactions (competitive, mutualistic, neutral\dots), the degree distribution, the number of directed loops of different lengths, the network spectrum, etc., remain near indistinguishable from their counterparts in random networks (see SI for details). This does not mean that more subtle differences (beside network sparsity) do not exist between both ensembles. As a matter of fact, such non-random structural features \emph{must} exist,  because typical representatives of the ensemble of random sparse networks do not comply with the empirical macroecological laws---only those found through the Monte Carlo simulations do. Identifying such additional non-random structural features is not a trivial task because it most likely involves deciphering higher-order correlations in the way pairwise interactions are placed within the network. Thus, it remains a challenging goal for future research.

An aspect that makes the problem especially challenging is the large number of free parameters (interactions) to be determined (it scales with the square of the number of species or taxa) as well as the intrinsic difficulty of identifying its actual values (even with perfect information on correlations; see \cite{Pinto2022}).

Our solution to such a formidable challenge is to construct not just one inferred network---specifying in a detailed way each possible pairwise interaction---but rather a whole ensemble of possible interaction networks compatible with the observed distribution of correlations. In this approach, the identity of specific pairs becomes irrelevant: it can neither be answered nor makes any difference when it comes to explaining macroecological data.

An alternative approach is that of Gibson et al. \cite{Gibson}. The authors lump microbial taxa into groups (or modules), assuming that all taxa within a group share the same interactions with other taxa outside their module (and respond to perturbations in the same way). This ``dimensionality reduction'' allows them to perform a (coarse-grained) network inference with a much smaller number of parameters. Let us stress, however, that we observed very little modularity in the matrices emerging out of our Monte Carlo method which rules out, in principle, such a coarse-grainning procedure.

We are aware that some modeling choices we have made can be questioned. For instance, the use of pairwise interactions in a Lotka-Volterra-like fashion. Apart from its long tradition in theoretical ecology, recent works \cite{Joseph2020,Dedrick2022} show that, within some limits, it is a reasonable choice. Nevertheless, it has been argued that higher-order interactions maybe crucial in the correct assessment of community stability and the understanding of its conflicting relationship with species diversity \cite{Bairey2016,Grilli2017}. Microbiomes are extraordinarily complex communities where processes involving more than two species may be prevalent \cite{Ludington2022}. As a matter of fact, effective dynamical models for species abundances derived from more detailed consumer-resource models, do generically include higher-order interactions (see e.g. \cite{Sireci2023}). Therefore, including higher-order interaction terms in \eqref{eq:mainEquation} is a generalization worth exploring.

Reproducing abundance correlations with direct interactions alone is more a proof of concept than an actual claim that this is the only mechanism for species correlations. We have already mentioned that phylogeny may be behind much of the positive correlation among microbial species \cite{Sireci2023} so that, in spite of the fact that environmental noise cannot reproduce by itself the observed correlation patterns, this does not mean that it must be ruled out. Most likely, both terms must be kept in \eqref{eq:mainEquation} and the truth lies in an appropriate combination of both. As a matter of fact, this might be the reason behind the asymmetry observed in the correlation patterns that the model with interactions alone struggles to capture.

Other important modeling ingredients, such as demographic noise or immigration, are left out as well. Demographic noise stems from intrinsic fluctuations in birth-and-death processes, and it is proportional to the square-root of the abundance (unlike the environmental noise of the SLVM, proportional to the abundance itself). A very recent study \cite{george2023} posits that a simple linear model with demographic noise is capable of reproducing patterns, not only of microbiomes, but of other very different systems as well. It has the additional advantage of being analytically solvable. The authors claim that such simple models can do a better job at capturing general properties observed across very different systems. Be that as it may, environmental noise can explain all three Grilli's laws---whereas, as argued in \cite{Grilli2020}, demographic noise cannot---and can also describe other statistical patterns (see Figs.~S21 and S22 of the SI). For these reasons, we have favored it over demographic noise in our model to study species-abundance correlations. Nevertheless, future work to further discriminate the effects and/or interplay of both types of noise would be highly desirable.

As for immigration, let us remark that it precludes extinctions in open ecosystems such as ours \cite{Hu2022}, but its effects are more relevant when communities are away from steady states. In this respect, recent work \cite{lim2023} emphasizes that the gut microbiome is constantly bottlenecked, with large amounts of biomass being constantly lost in the stool. It is clear then that abundances can change due to processes other than replication. Thus, extensions of our framework to account for these effect will need to be eventually incorporated.

Aside from these considerations, our analyses offer many possibilities to reach a deeper understanding of microbial communities and their emerging ecological patterns. For instance, whereas the origin of the gamma AFD---and its cousin, Taylor's law---is related to the multiplicative nature of the noise (the larger the abundance the larger its fluctuations), we still lack a good explanation for the appearance of a lognormal MAD. Both, in the SLM and the present SLVM it has been imposed by purposely tailoring the carrying capacities of the species. But through \eqref{eq:shifts}, the SLVM offers the possibility that a special choice of the interaction constants---away from what we have termed weak-interaction regime---may induce a lognormal MAD in some self-organized way. Preliminary analyses show that this can actually happen (see Sec.~6 of the SI), placing the explanatory burden on the nature of the interaction networks. This launches network theory of species interactions in the long-lasting debate \cite{McGill2007} about the origin of mechanistic processes behind the emergence of heavy-tailed species-abundance distributions---something that, to the best of our knowledge has been scarcely explored so far (see \cite{Wilson2003} for an exception).

Perhaps the most important message of the present work is that direct interactions between species are as relevant in microbiomes as they are in other more traditional ecosystems---such as animal-plant communities or food webs. In this regard, our analysis brings the study of microbes closer to the well-established framework of community ecology, where generalized Lotka-Volterra models play a central role. This paves the way to testing theoretical laws in ecology through experiments performed in microbial communities. The test of the stability-diversity relationship carried out in Ref.~\cite{Yonatan2022} is an excellent example of this idea. In view of the usual scarcity of data for traditional ecosystems, the overwhelming amount of microbial data provided by metagenomics opens an avenue of unprecedented possibilities for ecology.

\subsection*{Numerical solution of the SLVM}

Equation~(\ref{eq:mainEquation}) was solved numerically using an Euler-Maruyama integration scheme \cite{Toral2014}. For each species, the solution represents a noisy logistic trajectory, with the stationary mean population determined by the interaction properties. Using the resulting timeseries, the population of a given species in different samples may be recovered---once the dynamics has reached the stationary state---by either selecting abundances at different times (longitudinal data) or considering the abundances of different realizations at the same time (cross-sectional data). Both ways lead to identical results (i.e., the system is ergodic). Further details are discussed in Sec.~1 of the SI.  

\subsection*{Environmental noise matrix sampling} To produce a random, positive definite, symmetric matrix $\mathbf{W}$ we factor it as $\mathbf{W}=\mathbf{U}\boldsymbol{\Lambda}\mathbf{U}^T$, where $\mathbf{U}$ is an $S \times S$ orthogonal matrix ($\mathbf{UU}^T=\mathbf{U}^T\mathbf{U}=\mathbf{I}$) and $\boldsymbol{\Lambda}$ is a diagonal matrix whose diagonal elements are random, non-negative real numbers. The matrix $\mathbf{U}$ can be generated by randomly sampling from a Haar distribution (generated using the Python function \texttt{ortho\_group} from the SciPy package \cite{2020SciPy-NMeth}). The diagonal elements of $\Lambda$ have been drawn from different probability distributions, but all of them lead to similar results (see S5 of the SI for a full account).

\subsection*{Bayesian approach}

The posterior distribution of matrix $\mathbf{A}$, given the correlation distribution $\rho$, is obtained from Bayes's formula
\begin{equation}\label{eq:BayesFormula}
P(\mathbf{A}|\rho) = \frac{P(\rho|\mathbf{A})P(\mathbf{A})}{P(\rho)}.
\end{equation} 
In order to sample matrices $\mathbf{A}$ from the posterior distribution $P(\mathbf{A}|\rho)$ we apply a Metropolis-Hastings factor algorithm \cite{Toral2014}. This amounts to replacing  such samples by those of a purposely tailored Markov chain. In particular, at each step $n$ of the chain a pair of species $(i,j)$ is randomly selected and its corresponding interaction constant is modified as $a^{(n+1)}_{ij}=a^{(n)}_{ij}+\eta$, where $\eta\sim\mathcal{U}(-\epsilon,\epsilon)$. This change is accepted with probability $\min(1,H_n)$---otherwise rejected---where the Hasting factor (using Bayes's formula \eqref{eq:BayesFormula}) is obtained as 
\begin{equation*}
H_n=\frac{P(\rho|\mathbf{A}^{(n+1)})P(\mathbf{A}^{(n+1)})}{P(\rho|\mathbf{A}^{(n)})P(\mathbf{A}^{(n)})}.
\end{equation*}
The log-likelihood is computed (up to a trivial additive constant) as
\begin{align*}
    \log P(\rho|\mathbf{A})=&\,-\frac{1}{2\Delta_1^2}\sum_i
    \left[\log\left(\frac{\rho(x_i)}{\hat\rho(x_i)}\right)\right]^2
    -\frac{1}{2\Delta_2^2}|2-\gamma| \\
    &-\frac{1}{2\Delta_3^2}\sum_i
    \left[\log\left(\frac{\ell(\bar x_i)}{\hat\ell(\bar x_i)}\right)\right]^2,
\end{align*}
where $\Delta_1=2$, $\Delta_2=0.1$, $\Delta_3=0.3$ are weights chosen to ensure that all cost terms are comparable (we have verified that the Monte Carlo is rather robust to their precise values); $\rho(x)$ is the empirical distribution of Pearson's coefficients; $\hat\rho(x)$ is the one computed using matrix $\mathbf{A}$; $\ell(\bar x_i)$ is the rescaled MAD as obtained from simulations; and $\hat\ell(\bar x_i)$ is a standardized lognormal distribution. (The AFD needs not be imposed because it is hardly affected by interactions; see Fig.~\ref{fig:macroLawsPlot}.) As for the prior $P(\mathbf{A})$, we choose it to be zero if $\mathbf{A}$ leads to an unstable or unfeasible community, and constant otherwise. Finally, $\epsilon$ is selected so that the acceptance ratio at the start of the Markov chain is $\sim$30\%. Notice though that this acceptance ratio decreases as the Monte Carlo progresses.

\subsection*{Double normal distribution}

The nonzero fraction $f$ of non-diagonal elements of the interaction matrices $\mathbf{A}$ generated through Monte Carlo simulations follow a distribution that can be well described as the convex combination of two zero-mean normal distributions with standard deviations $\sigma_w$ (for ``wide'') and $\sigma_n$ (for ``narrow''). Typically, $\sigma_w$ is at least an order of magnitude larger than $\sigma_n$. Since it spans so different size scales, this distribution is better visualized in a log-log scale. As a function of $z=\log|x|$ a normal distribution has the expression
\begin{equation}
\log g(z,\sigma)=\frac{1}{2}\log(2/\pi\sigma^2)-\frac{e^{2z}}{2\sigma^2}+z.
\end{equation}
Thus, in log-log scale, the distribution of the absolute value of coefficients (signs are equally likely positive or negative) is given by
\begin{equation}
G(|a_{ij}|)=\rho g(|a_{ij}|,\sigma_B)+(1-\rho)g(|a_{ij}|,\sigma_S).
\end{equation}
Thus, $\rho$ can be interpreted as the fraction of nonzero matrix elements that are ``large''. Hence, an ``effective connectance'' of the interaction matrix can be estimated as $C_{\text{eff}}=f\rho$.

\begin{acknowledgments}
This work has been supported by (i) grants PGC2022-141802NB-I00 (BASIC), PID2021-128966NB-I00, and PID2020-113681GB-I00 funded by MCIN/AEI/10.13039/501100011033 and by ``ERDF A way of making Europe'', and (ii) Consejer{\'\i}a de Conocimiento, Investigaci{\'o}n Universidad, Junta de Andaluc{\'\i}a and Universidad de Granada under project B-FQM-366-UGR20 (ERDF). We also thank Jacopo Grilli for enlightening discussions.
\end{acknowledgments}

\bibliographystyle{apsrev4-2}
\bibliography{bibliography.bib}

\end{document}